\documentclass[11pt]{article}
\usepackage{amssymb,latexsym,amsmath}
\usepackage{amsfonts, graphics}
\newcommand{\R}{\mathbb R}

\def\open#1{\setbox0=\hbox{$#1$}
\baselineskip = 0pt
\vbox{\hbox{\hspace*{0.4 \wd0}\tiny $\circ$}\hbox{$#1$}}
\baselineskip = 11pt\!}
\def\fn{\open{f}}

\newcommand{\prfe}{\hspace*{\fill} $\Box$

\smallskip \noindent}

\def\be{\begin{equation}}
\def\ee{\end{equation}}
\def\bea{\begin{eqnarray}}
\def\eea{\end{eqnarray}}
\def\beas{\begin{eqnarray*}}
\def\eeas{\end{eqnarray*}}

\def\supp{\mathrm{supp}\,}

\begin{document}

\newtheorem{theorem}{Theorem}[section]
\renewcommand{\thetheorem}{\arabic{section}.\arabic{theorem}}
\newtheorem{definition}[theorem]{Definition}
\newtheorem{proposition}[theorem]{Proposition}
\newtheorem{example}[theorem]{Example}
\newtheorem{remark}[theorem]{Remark}
\newtheorem{cor}[theorem]{Corollary}
\newtheorem{lemma}[theorem]{Lemma}

\title{The asymptotic behaviour in Schwarzschild time
of Vlasov matter in spherically
symmetric gravitational collapse}

\author{H{\aa}kan Andr\'{e}asson\thanks{Support by the 
        Institut Mittag-Leffler (Djursholm, Sweden) is gratefully 
        acknowledged.}\\
        Mathematical Sciences\\ University of Gothenburg\\
        Mathematical Sciences\\ Chalmers University of Technology\\
        S-41296 G\"oteborg, Sweden\\
        email: hand@chalmers.se\\
        \ \\
        Gerhard Rein\\
        Mathematisches Institut der Universit\"at Bayreuth\\
        D-95440 Bayreuth, Germany\\
        email: gerhard.rein@uni-bayreuth.de}

\maketitle

\begin{abstract}
Given a static Schwarzschild spacetime of ADM mass $M$, 
it is well-known that no ingoing causal geodesic starting in the outer domain 
$r>2M$ will cross the event horizon $r=2M$ in finite Schwarzschild time. 
In the present paper we show that in gravitational collapse of Vlasov matter
this behaviour can be very different. We construct initial data for which a
black hole forms and all matter crosses the event horizon as 
Schwarzschild time goes 
to infinity, and we show that this is a necessary condition for geodesic 
completeness of the event horizon. 
In addition to a careful analysis of the asymptotic behaviour 
of the matter characteristics our proof requires a new argument for global existence 
of solutions to the spherically symmetric Einstein-Vlasov system in an outer domain, 
since our initial data have non-compact support in the radial momentum variable 
and previous methods break down. 
\end{abstract}

\section{Introduction}
\setcounter{equation}{0}

In a previous study \cite{AKR1} two classes of initial data for the 
spherically symmetric Einstein-Vlasov system were constructed which guarantee 
the formation of black holes. An additional argument 
to match the definition of a black hole in \cite{DR2} is given \cite{AKR2}. 
In the present paper we denote any of these initial data classes 
by $\mathcal{I}$.

The analysis in \cite{AKR1} is carried out in Schwarzschild
coordinates where the metric takes the form
\begin{equation}\label{metric}
ds^2=-e^{2\mu(t,r)}dt^2+e^{2\lambda(t,r)}dr^2+
r^2(d\theta^2+\sin^2\theta\,d\varphi^2).
\end{equation}
Here $t\in\R$ is the time coordinate, $r\in [0, \infty[$ is the
area radius, i.e., $4 \pi r^2$ is the area of the orbit of the
symmetry group $\mathrm{SO}(3)$ labeled by $r$, and the angles
$\theta\in[0, \pi]$ and $\varphi\in[0, 2\pi]$ parameterize these
orbits. The structure of the initial data $\mathcal{I}$ is such
that a possibly large fraction of its ADM mass $M$ is
necessarily located in the outer domain $r>2M.$ In \cite{AKR1} 
it was shown that solutions 
launched by such initial data have the following
property: there exist constants $\alpha, \beta > 0$ such that
spacetime is vacuum for
\begin{equation}\label{rgreater}
r \geq 2 M + \alpha e^{-\beta t},\; t\geq 0.
\end{equation}
Hence in this domain the metric equals the Schwarzschild
metric
\[
ds^2 = -\left(1-\frac{2 M}{r}\right)\, dt^2 
+ \left(1-\frac{2 M}{r}\right)^{-1}dr^2\,
+ r^2(d\theta^2 + \sin^2 \theta d\varphi^2),
\]
representing a black hole of mass $M$. The generator of 
the event horizon approaches the surface $r=2M$ asymptotically 
as Schwarzschild time goes to infinity, cf. \cite[Thm.~2.4]{AKR1}. 

Although (\ref{rgreater}) gives information about the asymptotic location of 
the matter it does not answer the question whether or not matter crosses 
the surface $r=2M.$ 
As a matter of fact, the inequality (\ref{rgreater}) is not sufficient 
to conclude that any matter initially in the region $r>2M$ ever crosses 
the surface $r=2M$ since matter can pile up at $r=2M.$ 
On the other hand it is known that not all matter can cross the 
surface $r=2M$ in finite Schwarzschild time. Indeed, if this were to happen
the Einstein equations would imply that the metric 
function $\lambda$ became infinite at $r=2 M$.
But according to \cite{RRS} this cannot happen for
the solutions considered in  \cite{AKR1}. 
It follows that on any finite time interval some matter 
must remain in the region $r>2M.$ 
The purpose of the present paper is to investigate the asymptotic 
behaviour of Vlasov matter in Schwarzschild time 
in the neighbourhood of the event horizon. Note that 
if matter crosses the surface $r=2M$ 
in finite time it also crosses the event horizon in finite time. 
Our main motivations are the following.
\begin{itemize}
\item 
In Proposition~\ref{necessary} it is shown that a necessary condition for 
completeness of the outgoing radial null geodesic
which generates the event horizon is that all matter crosses the surface $r=2M$
as Schwarzschild time goes to infinity. 
\item
In a static Schwarzschild spacetime of ADM mass $M$ 
no ingoing causal geodesic starting in the outer domain 
$r>2M$ will cross the event horizon $r=2M$ in finite Schwarzschild time. 
It is interesting to know if this remains true in evolutionary 
gravitational collapse. The result in the present paper shows that for
the initial data we construct the behaviour is indeed very different. 
\item 
In \cite[p.~13]{D1} some open problems about gravitational 
collapse are stated. For instance, for a scalar field it is known that
\begin{equation}\label{supmatter}
\sup_{H}r=2\sup_{H}m,
\end{equation}
but for other matter models this issue is open. Here $m$ is 
the quasi-local mass, $H$ is the event horizon, and $r$ the area radius. 
\item
\cite[Thm.~1.5]{DR3} relates the asymptotic behaviour 
of the matter at the event horizon to the question
of strong cosmic censorship, see also 
\cite[Question 15.3]{DR3}.
\item 
The asymptotic behaviour of matter in Schwarzschild time is  
directly related to what earth bound observers of gravitational 
collapse observe, which is not the case using other standard 
coordinates, e.g. Eddington-Finkelstein coordinates. 
If in Schwarzschild time all the matter crosses $r=2M$, 
an earth bound observer will "see" all 
matter eventually swallowed by the emerging black hole.
\item 
An important open problem is the question whether or 
not solutions can break down in finite Schwarzschild time. It is often 
conjectured that Schwarzschild coordinates are singularity avoiding and 
that in these coordinates solutions of the spherically symmetric 
Einstein-Vlasov system exist globally for general initial data. 
We expect that to understand the asymptotics of Vlasov matter 
in gravitational collapse
is going to be useful for understanding the global existence issue in general. 
\end{itemize}

In the present paper we construct a class $\mathcal{J}$ of initial data 
for which the results in \cite{AKR1} apply, and such that all the matter
asymptotically crosses the surface $r=2M$. In particular (\ref{supmatter}) 
holds for Vlasov matter for this class of initial data. 
The class $\mathcal{J}$ is different from the class $\mathcal{I}$
in that the support of the momentum variables is 
not compact. 
This is a technical condition needed for our method
of proof, but we believe that the conclusion holds 
for compactly supported data as well. 
Below, we always have in mind the radial momentum variable 
when we discuss compactly or non-compactly supported initial data. 
Our method of proof does imply that matter 
crosses the surface $r=2M$ also in the compactly supported case,
but we are not able to conclude 
that \textit{all matter} eventually crosses $r=2M$ in this case. 
We point out that the condition of non-compact support is required 
in some works in the cosmological case, cf. \cite{DR2} and \cite{SJ}. For 
compactly supported initial data the result in \cite{RRS}
guarantees that solutions exist as long as matter stays in a region 
$r\geq\epsilon>0.$ The proof in \cite{RRS} breaks down for non-compactly
supported data. 
But for applying the method in \cite{AKR1} it is crucial  
that solutions are global in an outer domain, and so we need 
to establish such a global existence result 
in the case of non-compactly supported data as well. 
Non-compactly supported data have 
been considered for other kinetic equations 
such as the Vlasov-Poisson system \cite{H} and the 
Vlasov-Maxwell system \cite{GS,Sch}. 
These methods do not directly apply in the case of the 
Einstein-Vlasov system, 
and we have not been able to find a result analogous to \cite{RRS} 
for non-compactly supported data. 
However, the initial data set $\mathcal{J}$ constructed below is 
such that the matter continues to move inward for all times. 
This crucial feature allows us to obtain the necessary global
existence proof for the corresponding, non-compactly supported 
data in an outer domain. 

The outline of the paper is as follows. In the next section we
introduce the Einstein-Vlasov system, recall the set up and 
the construction of the class of initial data in \cite{AKR1}, 
and formulate the main results of the present paper. 
Sections~4, 5, and 6 are devoted to their proofs. 

\section{Set up and main results}
\setcounter{equation}{0}

In this section we recall the Einstein-Vlasov system and the set up in \cite{AKR1} 
and formulate the main results. For more background on kinetic theory and 
the Einstein-Vlasov system we refer to \cite{An}. We consider the asymptotically flat 
spherically symmetric Einstein-Vlasov system. 
We use Schwarzschild coordinates $(t,r,\theta,\varphi)$ and parameterize 
the metric as in (\ref{metric}). 
Asymptotic flatness means that the metric quantities $\lambda$ and $\mu$
have to satisfy the boundary conditions
\begin{equation}\label{boundc}
   \lim_{r\to\infty}\lambda(t, r)=\lim_{r\to\infty}\mu(t, r)=0.
\end{equation}
Vlasov matter is a collisionless ensemble of particles which is described 
by a density function $f$ on phase space. In order to exploit the symmetry 
it is useful to introduce
non-canonical variables on momentum space and write $f=f(t,r,w,L)$. 
The variables $w\in ]-\infty,\infty[$ and $L\in [0,\infty[$ can be
thought of as the radial component of the momentum and the square of
the angular momentum respectively. 

The Vlasov equation is given by
\begin{equation} \label{vlasov}
\partial_{t}f+e^{\mu-\lambda}\frac{w}{E}\partial_{r}f
-\left(\lambda_{t}w+e^{\mu-\lambda}\mu_{r}E-
e^{\mu-\lambda}\frac{L}{r^3E}\right)\,\partial_{w}f=0,
\end{equation}
where
\[
E=E(r,w,L):=\sqrt{1+w^{2}+L/r^{2}}, 
\]
and where subscripts indicate partial derivatives.
The Einstein equations read
\begin{equation}\label{ein1}
   e^{-2\lambda}(2r\lambda_r-1)+1=8\pi r^2 \rho,
\end{equation}
\begin{equation} \label{ein2}
   e^{-2\lambda}(2r\mu_r+1)-1 = 8\pi r^2 p,
\end{equation}
\begin{equation} \label{ein3}
   \lambda_t = -4 \pi r j,
\end{equation}
and the matter quantities are given by
\begin{eqnarray}
\rho(t,r)
&=&
\frac{\pi}{r^{2}}
\int_{-\infty}^{\infty}\int_{0}^{\infty}Ef(t,r,w,L)\,dL\,dw,
\label{rho}\\
p(t,r)
&=&
\frac{\pi}{r^{2}}\int_{-\infty}^{\infty}\int_{0}^{\infty}
\frac{w^{2}}{E}f(t,r,w,L)\,dL\,dw,
\label{p}\\
j(t,r)
&=&
\frac{\pi}{r^{2}}
\int_{-\infty}^{\infty}\int_{0}^{\infty}w\,f(t,r,w,L)\,dL\,dw.
\label{j}
\end{eqnarray}
The equations (\ref{vlasov})--(\ref{j}) constitute the spherically symmetric 
Einstein-Vlasov system in Schwarzschild coordinates. 
For a detailed derivation of this system 
we refer to \cite{R}. 

As initial data we need to prescribe an initial distribution function
$\open{f}=\open{f}(r,w,L)\geq 0$ such that
\begin{equation}\label{notsinit}
   \int_0^r 4\pi\eta^2\open{\rho}\,(\eta)\,d\eta
   =4\pi^2 \int_0^r\int_{-\infty}^{\infty}\int_0^{\infty}
   E\open{f}(\eta,w,L)\,dL\,dw\,d\eta < \frac{r}{2}.
\end{equation}
Here we denote by $\open{\rho}$ the energy density induced by the initial
distribution function $\open{f}$. 
If in addition the initial data is $C^1$ we say that it is regular. 
In previous investigations the condition of compact support was included 
in the definition of regular data, but compact support in $w$ is not required 
in the present paper and is replaced by a suitable fall-off condition,
cf.\ (\ref{iddecay}) below. 
The Cauchy problem is well defined for regular initial data. We 
will restrict ourselves to a smaller class of regular initial data which guarantee 
the formation of black holes. Clearly, black holes do not form for any initial data, 
e.g., if the data is sufficiently small matter disperses and spacetime is 
geodesically complete, cf.\ \cite{RR}. 

Let us recall the set up and the properties 
of one of the initial data sets constructed in \cite{AKR1}. 
We fix $0<r_0<r_1$, 
and let $\gamma^+$ be the outgoing radial null geodesic
originating from $r=r_0$, i.e.,
\begin{equation}\label{gamma+}
\frac{d \gamma^+}{ds}(s)=e^{(\mu-\lambda)(s,\gamma^+(s))},\;\gamma^+(0)=r_0.
\end{equation}
We consider solutions of the spherically symmetric Einstein-Vlasov system 
({\ref{vlasov})--(\ref{j}) on the outer region
\begin{equation} \label{ddef}
D:=\{(t,r) \in [0,\infty[^2 \mid r \geq \gamma^+(t)\}.
\end{equation}
Note that characteristics of the Vlasov equation can pass
from the region $D$ into the region $\{r < \gamma^+(t)\}$
but not the other way around so that initial data
$\fn$ posed for $r>r_0$ completely determine the solution
on $D$. 

Let $M:=r_1/2$ be the total ADM mass and define 
the quasi-local mass by
\begin{equation}\label{m-def}
m(t,r)=M - 4\pi\int_r^\infty \rho(t,\eta)\,\eta^2 d\eta.
\end{equation}
Let $M_\mathrm{out}<M$ be given and such that 
\begin{equation}\label{8over9}
\frac{2(M-M_\mathrm{out})}{r_0}<\frac{8}{9}.
\end{equation}
Take $R_1>r_1$ such that 
\[
R_1-r_1<\frac{r_1-r_0}{6},
\]
and define
\[
R_0:=\frac{1}{2}(r_1+R_1).
\]
We require that all the matter
in the region $[r_0, \infty[$ is initially located in the strip $[R_0,R_1]$,
with $M_\mathrm{out}$ being the corresponding fraction of the ADM mass $M$, i.e.,
\[
\int_{r_0}^{\infty}4\pi r^2\open{\rho}\,(r)\,dr = 
\int_{R_0}^{R_1}4\pi r^2\open{\rho}\,(r)\,dr=M_\mathrm{out}.
\]
Furthermore, the remaining fraction $M-M_\mathrm{out}$ should be
initially located within the ball of area radius $r_0$, i.e.,
\[
\int_{0}^{r_0}4\pi r^2\open{\rho}\,(r)\,dr=M-M_\mathrm{out}.
\]
If one considers the Einstein-Vlasov system on the whole spacetime
$r\geq 0$, then the definition (\ref{m-def}) for the quasi-local mass
is equivalent to the more standard one, namely 
$m(t,r) = 4\pi\int_0^r \rho(t,\eta)\,\eta^2 d\eta$.
This is because the ADM mass $M=m(t,\infty)$ is conserved.
On the outer domain $D$ the definition (\ref{m-def}) is more suitable,
since it does not refer to the matter inside $\{r< \gamma^+(t)\}$
except for the fact that this matter is there and contributes to
the total mass.
 
The properties above concern the structure of the initial data 
in space. We also need to specify conditions on the momentum variables. 
Let $W_-<0$ and $L_1>0$ be given. 
In \cite{AKR1} we introduced the {\bf{general support 
condition}}: For all 
$(r,w,L) \in \supp \fn ,$ 
\[
r \in ]0,r_0] \cup [R_0,R_1],
\]
and if $r\in [R_0,R_1]$ then
\[
w \leq W_-,\ 0\leq L \leq L_1,
\]
and
\[
0\leq L <\frac{3L}{\eta}\,\open{m}(\eta) +\eta\,\open{m}(\eta),\ \eta\in [r_0,R_1].
\]
Here we use the notation $\open{m}$ when $\rho=\open{\rho}\,$ in (\ref{m-def}).  
In addition to the conditions above the initial data $\mathcal{I}$ in \cite{AKR1} 
were assumed to have compact support. 
If $W_-$ is sufficiently negative, a black hole of ADM mass $M$
forms and 
$\lim_{t\to \infty} \gamma^\ast (t) = 2 M$
for a certain radially outgoing null geodesic which
is the generator of the event horizon, cf.\
\cite[Thm.~2.4]{AKR1} and \cite[Sect.~4.3]{AKR2}.
The initial data we construct below do have the properties specified
above, but the support in the radial momentum variable $w$ is not compact. 
However, the proof in \cite{AKR1} goes through unchanged also for 
such data
\textit{provided} the solutions are global on the domain $D$. 
With respect to global existence in the domain $D$ the following holds. 
\begin{theorem}\label{geforncdata}
Let regular initial data $\fn$ be given with the properties specified above,
and such that the fall-off condition (\ref{iddecay}) is satisfied. Then 
the corresponding solutions of the spherically symmetric Einstein-Vlasov 
system (\ref{vlasov})--(\ref{j}) in the domain $D$ exist for all $t\geq 0$. 
\end{theorem}
We can now state the main result of the present paper. 
\begin{theorem}\label{docross2M}
There exists a class of regular initial data for the spherically
symmetric Einstein-Vlasov system such that the corresponding
solutions have the asymptotic property that
\begin{equation}\label{mequalsM}
\lim_{t\to\infty}m(t,2M)=\lim_{t\to\infty}m(t,\gamma^\ast(t))=M.
\end{equation}
\end{theorem}
As mentioned in the introduction the condition (\ref{mequalsM}) is a necessary condition 
for completeness of the generator $\gamma^*$ of the event horizon. 
We state this in a proposition.
\begin{proposition}\label{necessary}
A necessary condition for future completeness of the generator $\gamma^*$ 
of the event horizon is that (\ref{mequalsM}) holds.
\end{proposition}

\section{Proof of Theorem~\ref{geforncdata}}
\setcounter{equation}{0}

Compactly supported, regular initial data launch a local regular solution
which can be extended as long as the momentum support of the solution
can be controlled \cite{R,RR}. We do not give a complete proof for the corresponding
result for non-compactly supported data and restrict ourselves to establishing
the main a-priori bounds. To this end it is convenient to introduce the
Cartesian coordinates 
$x=r(\sin\theta \cos \varphi, \sin \theta \sin \varphi, \cos \theta)\in \R^3$
with corresponding momentum variable $v\in \R^3$ so that
\begin{equation}\label{cart}
w = \frac{x\cdot v}{r},\ L =|x\times v|^2,\ |v|^2 = w^2 + \frac{L}{r^2}.
\end{equation}
Here $\cdot$ denotes the Euclidean scalar product and $|v|$ the induced norm.
We denote by $(X,V)(s,t,x,v)$ the solution of the characteristic
system of the Vlasov equation, written in the variables $x$ and $v$,
\beas
\dot x 
&=& 
e^{(\mu -\lambda)(s,r)} \frac{v}{\sqrt{1+|v|^2}},\\
\dot v
&=&
- \left(\lambda_t(s,r)\,\frac{x\cdot v}{r}  
+  e^{(\mu -\lambda)(s,x)} \mu_r(s,r)\,\sqrt{1+|v|^2}\right)\frac{x}{r},
\eeas
with $(X,V)(t,t,x,v) =(x,v)$; here $\dot {\phantom x}$ denotes the derivative
with respect to $s$.
We define
\begin{eqnarray}\label{Qsupp}
Q(t)
&:=&
\sup\left\{\frac{1+|v|}{1+|V(0,s,x,v)|} \mid 0\leq s\leq t,\ 
(x,v)\in \supp f(s) \right\}\nonumber\\
&=&
\sup\left\{\frac{1+|V(s,0,x,v)|}{1+|v|} \mid 0\leq s\leq t,\ 
(x,v)\in \supp \fn \,\right\}.
\end{eqnarray}
We require that the initial data satisfy the fall-off condition
\begin{equation}\label{iddecay}
\|\fn\,\|:=\sup_{(x,v)\in\R^6} (1+|v|)^5 |\fn (x,v)| < \infty.
\end{equation}
Since
\[
f(t,x,v)=\open{f}((X,V)(0,t,x,v)),
\]
we get the estimate
\[
f(t,x,v) \leq \|\open{f}\,\|(1+|V(0,t,x,v)|)^{-5}\leq Q^5(t)\|\open{f}\,\|(1+|v|)^{-5}.
\]
Hence
\begin{equation}\label{rhoest}
\int_{\mathbb{R}^3}(1+|v|)f(t,x,v)dv\leq 
\|\open{f}\,\|Q^{5}(t)\int_{\mathbb{R}^3}(1+|v|)^{-4}dv
\leq C\|\open{f}\,\|Q^5(t).
\end{equation}
By the characteristic system,
\[
\frac{d}{ds}(1+|V(s,0,x,v)|)
\leq (\|e^{\mu-\lambda} \mu_r(s) \|_{\infty}+\|\lambda_t (s)\|_{\infty})(1+|V(s)|).
\]
This implies that for $0\leq s\leq t$,
\[
\frac{1+|V(s,0,x,v)|}{1+|v|}
\leq e^{\int_0^t(\|e^{\mu-\lambda} \mu_r(\tau)\|_{\infty}+\|\lambda_t (\tau)\|_{\infty})d\tau},
\]
and we obtain the estimate
\[
Q(t)\leq e^{\int_0^t(\|e^{\mu-\lambda} \mu_r(s) \|_{\infty}+\|\lambda_t (s)\|_{\infty})ds}.
\]
The field equations (\ref{ein1}) and (\ref{ein2}) together with the boundary
condition (\ref{boundc}) imply that
\[
(\mu + \lambda)(t,r) = -\int_r^\infty (\mu_r + \lambda_r)(t,\eta)\,d\eta \leq
0,
\]
and 
\[
e^{\mu-\lambda} \mu_r(t,r) = e^{\mu+\lambda}
\left(\frac{m(t,r)}{r^2} + 4 \pi r p(t,r)\right)
\leq 4 \pi r \left(||\rho(t)||_\infty + ||p(t)||_\infty\right).
\]
Together with (\ref{ein3}) and (\ref{rhoest}) we have
\[
\|e^{\mu-\lambda}\mu_r (s) \|_{\infty}+\|\lambda_t (s)\|_{\infty}
\leq C(1+s)\|\open{f}\,\|Q^5(s),
\]
so that
\[
Q(t)\leq e^{\int_0^t C\|\open{f}\,\|(1+s)Q^5(s)ds}.
\]
This implies that $Q$ is bounded on some time interval $[0,T[$.
A standard iterative procedure then shows that there is a local,
regular solution which can be extended as long as the function
$Q$ does not blow up, cf.\ \cite{R,RR}.

Global existence in the outer domain $D$ will now follow if we can
establish a bound on $Q(t)$ in $D.$
For this argument we use the variables $(r,w,L)$.
By (\ref{iddecay}),
\[
\open{f}(r,w,L)\leq C|w|^{-3}.
\]
Let us define a quantity as in (\ref{Qsupp}). By abuse of notation
we let
\begin{eqnarray}\label{Qwsupp}
Q(t)
&:=&
\sup\left\{\frac{|w|}{|W(0,s,r,w,L)|} \mid
0\leq s\leq t,\;(r,w,L)\in \supp f(s) \right\}\nonumber\\
&=&
\sup\left\{\frac{|W(s,0,r,w,L)|}{|w|} \mid 0\leq s\leq t,\;
(r,w,L)\in \supp \open{f}\, \right\};
\end{eqnarray}
notice that in the outer domain $D$ the area radius $r\geq r_0>0$ 
so that by (\ref{cart}) a bound on $Q$ as defined in (\ref{Qwsupp})
implies a bound on $Q$ as defined in (\ref{Qsupp}).
The following lemma taken from \cite{AKR1} shows that 
when the general support condition holds,
then the particles in the outer domain $D$ keep moving inward
in a controlled way. 
\begin{lemma}\label{ingoinglemma}
Let $\fn$ be regular and satisfy the general support condition
for some  suitable $W_-<0$.
Then for all characteristics $(R(t),W(t),L)$ with $(R(0),W(0),L)\in \supp \fn$
and $R(0) \in [R_0,R_1]$,
\[
W(t) \leq e^{\lambda(t,R(t))} e^{-\lambda(0,R(0))} W(0) \leq
e^{-\lambda(0,R(0))} W(0)
\]
as long as $(t,R(t))\in D$.
In particular $w < 0$ for all
$(r,w,L) \in \supp f(t)$ and $(t,r) \in D$, and $j \leq 0$ on $D$.
\end{lemma}
Following \cite{RRS} we find that along any characteristic in $\supp f$,
\begin{equation}\label{charwsquared}
\frac{d}{ds}w^2
\leq Cw^2+C\int_{-\infty}^{\infty}\int_{0}^{L_1}|\tilde{w}|f(s,r,\tilde{w},\tilde L)
\,d\tilde L\,d\tilde{w};
\end{equation}
for this estimate it is essential that all particles are moving inward.
We estimate the last term.
Since 
\[
f(s,r,w,L)=\open{f}(R(0,s,r,w,L),W(0,s,r,w,L),L),
\]
we have
\begin{eqnarray*}
& &
\int_{-\infty}^{\infty}\int_{0}^{L_1}|\tilde{w}|f(s,r,\tilde{w},\tilde{L})
\,d\tilde L\,d \tilde{w}\nonumber\\
& &
\qquad \qquad \leq C\,\iint_{\supp{f(s,r,\cdot,\cdot)}}|\tilde{w}|
|W(0,s,r,\tilde{w},\tilde{L})|^{-3}\,d\tilde L\,d \tilde{w}.
\end{eqnarray*}
By the definition of $Q$ and the general support condition,
\[
|W(0,s,r,\tilde{w},\tilde{L})|\geq \max \left\{|W_-|,\frac{|\tilde{w}|}{Q(s)}\right\}.
\]
Hence we find that
\begin{eqnarray*}
\int_{-\infty}^{\infty}\int_{0}^{L_1}|\tilde{w}|f(s,r,\tilde{w},\tilde{L})
\,d\tilde L\,d\tilde{w}
&\leq& 
C \int_0^{\infty}\!\!\int_{0}^{L_1}\tilde{w} 
\left(\max\left\{|W_-|,\frac{\tilde{w}}{Q(s)}\right\}\right)^{-3}d\tilde{w}\\
&\leq& 
C \int_0^{|W_-|Q(s)}\tilde{w}\,d\tilde{w}+
\int_{|W_-| Q(s)}^{\infty}\tilde{w}\,\frac{Q^3(s)}{\tilde{w}^3}d\tilde{w}\\ 
&\leq& 
C\,Q^2(s).
\end{eqnarray*}
We have thus derived the estimate
\[
\frac{d}{ds}w^2\leq C\,w^2+C\,Q^2(s),
\]
and hence for $0\leq s\leq t$,
\[
\frac{w^2(s)}{w^2(0)}\leq e^{Ct}\left(1+C \int_0^t Q^2(\tau)\,d\tau\right).
\]
This implies that
\[
Q^2(t)\leq e^{C t}\left(1+C \int_0^t Q^2(s)\,ds\right),
\]
hence $Q$ is bounded on bounded time intervals, 
and global existence in $D$ follows.
\prfe

\section{Proof of Theorem~\ref{docross2M}}
\setcounter{equation}{0}

We aim to show that all characteristics starting in the domain $[R_0,R_1[$ 
enter the region $\{r \leq 2 M\}$ in finite time,
and we need an estimate for the required time. 
Let $(R(s),W(s),L)$ be a characteristic emanating from
the support of $\fn$ with $R(0) \in [R_0,R_1]$; all the following 
estimates are valid as long as $R(s) \geq 2 M$.
By Lemma~\ref{ingoinglemma} and the characteristic equation,
\[
\dot R(s)
=
\frac{W(s)}{E(s)}e^{(\mu-\lambda)(s,R(s))}
\leq
-B(R(0),W(0)) e^{(\mu-\lambda)(s,R(s))},
\]
where
\[
B(r,w) 
:= \frac{e^{-\lambda(0,r)}|w|}
{\sqrt{1+e^{-2 \lambda(0,r)} w^2 + L_1 (2M)^{-2}}}.
\]
We require that on $\supp \fn$, 
\begin{equation} \label{wcond}
w \leq - e^{\lambda(0,r)} K(r)
\end{equation}
where $K : [R_0,R_1[ \to ]0,\infty[$ is an increasing function
which will be specified below.
Hence
\[
B(r,w) \geq \frac{K(r)}{\sqrt{1+K^2(r) + L_1 (2M)^{-2}}} := B(r),
\]
and
\begin{equation} \label{dotr0}
\dot R (s)
\leq
-B(R(0))\,e^{(\mu-\lambda)(s,R(s))}.
\end{equation}
By \cite[Lemma 4.1 (a)]{AKR1},
\[
\mu - \lambda \geq 2 \hat\mu
\]
where
\[
\hat\mu(t,r) := - \int_r^\infty \frac{m(t,\eta)}{\eta^2} e^{2\lambda(t,\eta)}
d\eta .
\]
Inserting this into (\ref{dotr0}) implies that
\begin{equation} \label{dotr}
\dot R (s)
\leq
-B(R(0))\, e^{2 \hat\mu (s,R(s))}.
\end{equation}
In order to estimate the right hand side we compute
$\hat{\mu}_t$, cf.\ \cite[Lemma 4.1 (d)]{AKR1}, and
observe that
\begin{eqnarray*}
\hat{\mu}_t(s, r)
&=&
\int_r^\infty 4\pi j(s, \eta)
\,e^{(\mu+\lambda)(s,\,\eta)}e^{2\lambda(s,\,\eta)}\,d\eta \\
&\geq&
\frac{1}{2 r}
\int_r^\infty 4\pi\eta\,2 j(s, \eta)
\,e^{(\mu+\lambda)(s,\,\eta)}e^{2\lambda(s,\,\eta)}\,d\eta;
\end{eqnarray*}
note that by Lemma~\ref{ingoinglemma}, $j \leq 0$. Since 
\[
E +2 w +\frac{w^2}{E}
=\left(\sqrt{E} +  \frac{w}{\sqrt{E}}\right)^2\geq 0
\]
the expressions for the matter terms imply that
$2 j \geq -(\rho+p)$ so that by \cite[Lemma 4.2]{AKR1},
\begin{eqnarray*}
\hat{\mu}_t(s, r)
&\geq&
- \frac{1}{2 r}  
\int_r^\infty 4\pi\eta\,(\rho+p)(s, \eta)
\,e^{(\mu+\lambda)(s,\,\eta)}e^{2\lambda(s,\,\eta)}\,d\eta \\
&=&
-\frac{1}{2 r}\Big(1-e^{(\mu+\lambda)(s,\,r)}\Big).
\end{eqnarray*}
Moreover,
\[
\hat{\mu}_r(s, r) = \frac{m(s,r)}{r^2} e^{2\,\lambda(s,r)}
\]
and $|\dot R(s)| \leq e^{(\mu-\lambda)(s,R(s))}$. Hence
\begin{eqnarray}\label{est}
&&
\hat{\mu}(t,R(t))-\hat{\mu}(0,R(0))
=
\int_{0}^t \frac{d}{ds}\,\hat{\mu}(s,R(s))\,ds\nonumber \\
&&
\qquad = 
\int_{0}^t\Big(\hat{\mu}_t(s,R(s))
+\hat{\mu}_r(s,R(s))\dot{R}(s)\Big)\,ds\nonumber \\
&&
\qquad\geq 
\int_{0}^t\left(-\frac{1}{2R(s)}\Big(1-e^{(\mu+\lambda)(s,R(s))}\Big)
-\frac{m(s,R(s))}{R(s)^2}\,e^{(\mu+\lambda)(s,R(s))}\right)\,ds\nonumber\\
&& 
\qquad = 
\int_{0}^t\left( -\frac{1}{2R(s)}+\Big(\frac{1}{2R(s)}
-\frac{m(s,R(s))}{R(s)^2}\Big)\,e^{(\mu+\lambda)(s,R(s))}\right)\,ds.
\end{eqnarray}
The right hand side of this inequality will be estimated using
the following lemma.
\begin{lemma}\label{monotonicity}
Let $\rho \in L^1([2 M,R_1])$ be such that $\rho \geq 0$ and
$0\leq 2 m(r)/r<1$, where
\[
m(r) := M-\int_r^{R_1}4\pi \eta^2 \rho(\eta)\,d\eta,
\]
and let
\[
e^{2 \lambda(r)} := \left(1-\frac{2 m(r)}{r}\right)^{-1},\ r\in [2 M,R_1].
\]
Then for all $r\in [2 M, R_1]$,
\[
e^{-2 \int_r^{R_1}4\pi \eta\, \rho(\eta)\, e^{2 \lambda(\eta)}d\eta}
\geq\frac{r-2M}{r-2m(r)}.
\]
\end{lemma}
\textit{Proof.}
For $r\in [2 M, R_1]$ we define
\[
h(r) := (r-2 m(r))\, e^{g(r)}- r + 2 M,\quad
g(r) := -2 \int_r^{R_1}4\pi \eta \rho(\eta) e^{2 \lambda(\eta)}d\eta.
\]
Then
\beas
h'(r)
&=&
(1-8 \pi r^2 \rho(r))\, e^{g(r)} + (r-2 m(r))\, e^{g(r)} g'(r) - 1\\
&=&
(1-8 \pi r^2 \rho(r))\, e^{g(r)} + 
r\, \left(1-\frac{2 m(r)}{r}\right)\, e^{g(r)}
8 \pi r \rho(r) e^{2 \lambda(r)} -1\\
&=&
(1-8 \pi r^2 \rho(r))\, e^{g(r)} + r e^{-2 \lambda(r)} e^{g(r)}
8 \pi r \rho(r)\, e^{2 \lambda(r)} -1\\
&=&
e^{g(r)} -1 \leq 0,\ r\in [2 M, R_1].
\eeas
Hence for $r\in [2 M, R_1]$,
\[
h(r) \geq h(R_1) = 0,
\]
which is the assertion. \prfe
 
\smallskip

\noindent
\textbf{Remark. }It is interesting to note that the configuration for which 
equality holds in the inequality in the lemma can be shown to be an infinitely 
thin shell. This should be compared to the situation considered in 
\cite{A2} where an infinitely thin shell is the maximizer of a 
similar integral expression as above. 

\smallskip

\noindent
Let us return to the proof of Theorem~\ref{docross2M}. 
By the field equations (\ref{ein1}) and (\ref{ein2})
and the form (\ref{rho}) and (\ref{p}) of the matter terms,
\[
\mu_r + \lambda_r = 4 \pi r e^{2 \lambda} (\rho + p)
\leq 8 \pi r e^{2 \lambda} \rho .
\]
Hence, Lemma~\ref{monotonicity} implies that for $r\in [2 M, R_1]$,
\[
e^{(\mu+\lambda)(s,r)}\geq e^{-2 \int_r^{R_1}4\pi \eta\, \rho(s,\eta)\, 
e^{2 \lambda(s,\eta)}d\eta}\geq
\frac{r-2M}{r-2m(s,r)}.
\]
We insert this into the estimate (\ref{est}) and find that
as long as $R(t)\geq 2M$,
\[
\hat{\mu}(t,R(t)) \geq \hat{\mu}(0,R(0))-\int_{0}^t
\frac{M}{R^2(s)}ds.
\]
By (\ref{dotr}) this implies that
\[
\dot R (s)
\leq
-B(R(0))\, C(R(0)) \exp\left(-2 \int_{0}^s \frac{M}{R^2(\tau)}d\tau \right),
\]
where $C(r):= e^{2 \hat \mu (0,r)}$. This implies that
\beas
\frac{d}{ds}\frac{1}{R(s)}
&=& 
- \frac{\dot R(s)}{R^2(s)} \\
&\geq&
\frac{B C}{R^2(s)}\exp\left(-2 \int_{0}^s \frac{M}{R^2(\tau)}d\tau \right)
=
- \frac{B C}{2 M} 
\frac{d}{ds}\exp\left(-2 \int_{0}^s \frac{M}{R^2(\tau)}d\tau \right)
\eeas
which upon integration yields the estimate
\bea
\frac{1}{R(t)} 
&\geq&
\frac{1}{R(0)}
- \frac{B C}{2 M}
\left(\exp\left(-2 \int_{0}^t \frac{M}{R^2(s)}ds \right)-1\right) \nonumber\\
&\geq&
\frac{1}{R(0)}
+ \frac{(B C)(R(0))}{2 M}
\left(1-e^{- 2 M t /R_1^{2}}\right).\label{oneoverrest}
\eea
In order to proceed the functions $C$ and $B$
must be related properly. We require that
\[
\open{\rho}\,(r) > 0\ \mbox{for}\ r\in ]R_0,R_1[
\]
so that $\open{m}(r)< M$ for $r\in[0,R_1[$, and 
\beas
C(r) 
&=& 
e^{2\hat{\mu}(0,r)}
=
\exp\left(-\int_r^{\infty}\frac{2 \open{m}(\eta)}{\eta^2(1-2 \open{m}(\eta)/\eta)}d\eta\right)\\
&>&
\exp\left(-\int_r^{\infty}\frac{2M}{\eta (\eta-2M)}d\eta\right)=1-\frac{2M}{r}.
\eeas
We can therefore choose the function $K$ which specifies
our support condition (\ref{wcond})
in such a way that for $r\in [R_0,R_1[$,
\begin{equation}\label{Kcondition}
B(r)\; C(r) = \frac{K(r)}{\sqrt{1+K^2(r)+L_1(2M)^{-2}}}\;C(r) > 1-\frac{2M}{r};
\end{equation}
note that this necessarily implies that $K(r)\to\infty$ as $r\to R_1$,
and that the function $C$ is determined by the initial data.
Given any $r^\ast \in ]R_0,R_1[$ there now exists $\kappa >0$
such that for 
$r\in [R_0,r^\ast]$,
\[
B(r)\; C(r) > 1-\frac{2M}{r} + \kappa.
\]
The estimate (\ref{oneoverrest}) therefore implies that for any
characteristic as above, but with $R(0)\in [R_0,r^\ast]$, and as long
as $R(t)\geq2 M$,
\beas
\frac{1}{R(t)}
&\geq&
\frac{1}{R(0)}
+ \frac{1}{2 M}\left(1-\frac{2M}{R(0)} + \kappa\right)
\left(1-e^{- 2 M t /R_1^{2}}\right)\\
&=&
\frac{1+\kappa}{2M}\left(1-e^{- 2 M t /R_1^{2}}\right)
+ \frac{1}{R(0)}e^{- 2 M t /R_1^{2}}\\
&>&
\frac{1+\kappa}{2M}\left(1-e^{- 2 M t /R_1^{2}}\right),
\eeas
which implies that
\[
R(t) < \frac{2M}{1+\kappa}\left(1-e^{- 2 M t /R_1^{2}}\right)^{-1}.
\]
This shows that there is a time $t^\ast >0$ such that
$R(t^\ast) \leq 2 M$
for all characteristics starting with  $R(0)\in [R_0,r^\ast]$.

To complete the proof we
fix $\epsilon>0$ and let $r_{\epsilon}<R_1$ be sufficiently close to $R_1$ 
such that $\open{m}(r_{\epsilon})\geq M-\epsilon$. 
Then there is a finite time 
$t_{\epsilon}$ such that all characteristics $(R(t),W(t),L)$ with 
$R(0)\in [R_0,r_{\epsilon}]$ reach $r=2M$ at some time
$t\leq t_{\epsilon}$. 

We construct a curve $(t,\alpha(t))$ with 
the property that $m(t,\alpha(t))\geq M-2\epsilon$ for $0\leq t\leq t_{\epsilon}$ 
and  $\alpha(t)=2M$ for some $t\leq t_{\epsilon}$. 
To this end, let
\begin{equation}\label{delta}
\delta:=\frac{\epsilon}{4\pi R_1^2 t_\epsilon},
\end{equation}
and let $\alpha$ be the solution of 
\begin{equation}\label{alpha}
\dot\alpha = \frac{j(s,\alpha)-\delta}{\rho(s,\alpha)+\delta}
\,e^{(\mu-\lambda)(s,\alpha)},
\ \alpha(0)=r_{\epsilon}.
\end{equation}
The reason for introducing the $\delta$ parameter is to avoid any potential 
problems with uniqueness of solutions if $\rho=0$. 
Taking the partial derivative of $e^{-2\lambda}=1-2m/r$ 
with respect to $t$ and using (\ref{ein3}) we find that
$m_t = - 4 \pi r^2 e^{\mu -\lambda} j$, and hence
\beas
\frac{d}{dt}m(t,\alpha(t))
&=&
-4\pi e^{(\mu-\lambda)(t,\alpha(t))} \alpha^2(t) j(t,\alpha(t))+
4\pi\alpha^2(t)\rho(t,\alpha(t))\dot \alpha(t)\\
&=& 
4\pi \alpha^2 e^{(\mu-\lambda)(t,\alpha(t))}
\left(\frac{j-\delta}{\rho +\delta} \rho - j\right)\\
&\geq&
- 4\pi \alpha^2(t) e^{(\mu-\lambda)(t,\alpha(t))}\delta
\geq
- \frac{\epsilon}{t_{\epsilon}}.
\eeas
Here we used that $j\leq 0.$
Thus for all $0\leq t\leq t_\epsilon$,
\[
m(t,\alpha(t))\geq M-2\epsilon,
\]
and it remains to show that $\alpha (t) \leq 2 M$
for some $0 < t\leq t_\epsilon$ .
Define for each $r\in [R_0,R_1]$ the barrier curve
$(t,R_B(t))$ by
\[
\dot R_B = -B(r)e^{(\mu-\lambda)(s,R_B)},\ R_B (0) = r.
\]
We use the term barrier curve since the area 
radius along this curve is larger than the area radius along any characteristic 
$(R(t),W(t),L)$ starting in the support of $\fn$ with $R(0)=r$. This is 
clear from the differential estimate (\ref{dotr0}) 
and the estimates which followed.
In addition, all barrier curves starting at some $r\in [R_0,r_\epsilon]$
reach the region $r\leq 2 M$ within the time interval $[0,t_\epsilon]$.
For $r\in [R_0,R_1]$ the definition of $B(r)$ and the condition
on the support of $\fn$ imply that,
\beas
&&
|j(0,r)|
\geq
e^{\lambda(0,r)} B(r) \frac{\pi}{r^2} 
\int_{-\infty}^\infty \int_0^\infty 
\sqrt{1+e^{-2 \lambda(0,r)} w^2 + L_1 (2M)^{-2}}\fn\, dL\, dw\\
&&
\qquad
\geq B(r) \rho(0,r),
\eeas
and since $B<1$ we thus have 
\[
\frac{|j(0,r)|+\delta}{\rho(0,r)+\delta} > B(r). 
\]
Consider the barrier curve $R_B(t)$ with $R_B(0)=r_\epsilon$.
Then
\[
\dot \alpha (0) = \frac{j(0,r_\epsilon)-\delta}{\rho(0,r_\epsilon)+\delta}
e^{(\mu-\lambda)(0,r_\epsilon)}
< - B(r_\epsilon)\, e^{(\mu-\lambda)(0,r_\epsilon)} 
=\dot R_B(0),
\]
and hence $\alpha(s)<R_B(s)$ on a time interval $]0,s_1]$. 
Assume that $\alpha (t) > 2 M$ for all $0 \leq t \leq t_\epsilon$. 
We define
\[
t^\ast := \inf\left\{t\in [0,t_\epsilon] 
\mid
R_B < \alpha\ \mbox{on}\ [t,t_\epsilon]\
\mbox{for all barriers}\
\mbox{starting in}\ [R_0,r_\epsilon]\right\}.
\] 
Then there exists some barrier curve $R_B(t)$ starting
at some $r\in [R_0,r_\epsilon]$ such that 
$\alpha(t^\ast) = r^\ast := R_B(t^\ast)$. By definition of $t^\ast$,
$\alpha(t) > R_B(t)$ for $t>t^\ast$, and hence
\[
-\frac{|j(t^\ast,r^\ast)|+\delta}{\rho(t^\ast,r^\ast)+\delta}
e^{(\mu-\lambda)(t^\ast,r^\ast)} = \dot \alpha(t^\ast)
\geq
\dot R_B(t^\ast)
= - B(R_B(0)) e^{(\mu-\lambda)(t^\ast,r^\ast)},
\]
which implies that
\[
|j(t^\ast,r^\ast)| < B(R_B(0))\, \rho(t^\ast,r^\ast).
\]
The latter inequality is only possible if there is at least
one characteristic $(R(t),W(t),L)$ with $R(t^\ast)=r^\ast$,
and 
\beas
&&
\frac{K(R_B(0))}{\sqrt{1+K^2(R_B(0)) + L_1 (2M)^{-2}}} = B(R_B(0))
> \frac{|W(t^\ast)|}{\sqrt{1+|W(t^\ast)|^2 + L R(t^\ast)^{-2}}} \\
&&
\qquad
\geq B(R(0),W(0)) \geq B(R(0))
= \frac{K(R(0))}{\sqrt{1+K^2(R(0)) + L_1 (2M)^{-2}}}.
\eeas
Since the function $K$ is taken to be increasing this estimate
implies that $R(0) < R_B(0) \leq r_\epsilon$.
The barrier curve $(t,\tilde{R}_B(t))$ with $\tilde{R}_B(0)=R(0),$ 
must satisfy the estimate
$\alpha(t^\ast) = r^\ast = R(t^\ast) < \tilde{R}_B(t^\ast)$, 
and this is a contradiction to the definition of $t^\ast$.
Hence $\alpha (t) \leq 2 M$ for some $0 < t\leq t_\epsilon$, 
which proves that $\lim_{t\to \infty} m(t,2M) = M$.

We can chose the parameter $\delta$ in (\ref{delta})
such that $m(t,\alpha(t)) \geq M - 2\epsilon$
for all $0\leq t \leq t_\epsilon + 1$.
Since $\alpha$ is strictly decreasing, $\alpha(t_\epsilon +1) < 2 M$,
and since $\lim_{t\to \infty} \gamma^\ast(t) = 2 M$
there exists some time $t \geq t_\epsilon  +1$ such that
$\gamma^\ast (t) \geq \alpha(t_\epsilon +1)$ and hence
$m(t,\gamma^\ast (t)) \geq M - 2\epsilon$.
This completes the proof of Theorem~\ref{docross2M}. \prfe

To conclude our main result  
we show that initial data
which satisfy the conditions required above do exist.
To this end, let $\rho=\rho(r)$ be a
$C^1$ function supported in $[R_0,R_1]$ with 
$\rho(r)>0$ on $]R_0,R_1[,$ and such that 
\[
M_{\mathrm{out}}=4 \pi \int_{R_0}^{R_1}r^2\rho(r)dr<M,
\]
satisfies (\ref{8over9}).
Define
\[
m(r):=M - 4\pi \int_{r}^{\infty}r^2\rho(r)dr,\ 
e^{-2 \lambda(r)} := 1 - \frac{2 m(r)}{r}. 
\]
If $r\in [R_0,R_1[$, then $M-m(r)>0$, and hence there exists a function
$K$ as introduced in (\ref{wcond}), which satisfies the condition
(\ref{Kcondition}).
Now let $\tilde{h}=\tilde{h}(r,w,L)$ be a $C^1$
function supported in $[R_0,R_1]\times ]-\infty,\infty[\times [0,L_1]$ and
such that $\tilde{h}(r,w,L)=0$ if $e^{- \lambda(r)}w>-K(r)$, and
\[
\int_{-\infty}^{\infty}\int_0^{L_1}\tilde{h}(r,w,L)\,dL\,dw = \frac{r^2}{\pi}.
\]
Let 
\[
h(r,w,L)=\frac{\tilde{h}(r,w,L)}{\sqrt{1+w^2+L/r^2}}
\]
so that $\rho(r) h(r,w,L)$ induces the energy density
$\rho$ and the quasi-local mass $m$.
Let $f_i$ be a density
function supported in $[0,r_0]$ such that the
assumptions of Lemma~\ref{ingoinglemma} hold. 
Then $\fn =f_i+\rho\,h$
defines initial data which have all the required properties
stated above.

\section{Proof of Proposition \ref{necessary}}
\setcounter{equation}{0}

The function $m(t,r)$ is increasing in both variables. 
We assume that
$m(t,r) \leq m(t,2 M) \leq C < M$ for all $t\geq 0$ and $r\leq 2 M$
and have to show that the generator $\gamma^*$ of the event horizon
is incomplete.
According to \cite[Thm.~2.4]{AKR1}, $\gamma^*$ approaches $r=2M$ as $t\to\infty$. It
follows that for all $t\geq 0$,
\begin{equation}\label{lambdabound}
\lambda(t,\gamma^*(t))\leq C.
\end{equation}
Let $\tau \mapsto (t,r,\theta,\phi)(\tau)$ be an affine parameterization
of $\gamma^*$ with corresponding momenta $(p^0,p^1,p^2,p^3)(\tau)$.
Since $\gamma^*$ is radial, let $\theta=\pi/2,\ \phi=0$
and $p^2 = p^3 = 0$. Since $\gamma^*$ is null, 
$$e^{2\mu}(p^0)^2 = e^{2\lambda} (p^1)^2,$$
and we get
\begin{equation}\label{p0asp1}
p^0=e^{\lambda-\mu}p^1.
\end{equation}
By the geodesic equations,
\begin{eqnarray*}
\frac{d p^1}{d\tau} 
&=&
 -e^{2(\mu - \lambda)} \mu_r (p^0)^2 - \lambda_r (p^1)^2
-2 \lambda_t p^0 p^1\\
&=&
4\pi \gamma^* e^{2\lambda}(p^1)^2[-p-\rho+2j].
\end{eqnarray*}
Here we used (\ref{p0asp1}) to express $p^0$ in terms of $p^1.$
Since $dt/d\tau =p^0$,
\begin{equation}\label{dp1dt}
\frac{dp^1}{dt}=4\pi \gamma^* e^{\mu+\lambda}p^1[-(p+\rho)+2j],
\end{equation}
and $\gamma^*$ is incomplete if
\[
\int_0^{\infty}\frac{ds}{p^0}<\infty.
\] 
By (\ref{p0asp1}) and  (\ref{dp1dt}),
\begin{equation}\label{intt}
\int_0^t\frac{ds}{p^0}=\int_0^t\frac{e^{\mu-\lambda}}{p^1}ds
=\int_0^t e^{\mu-\lambda}\frac{1}{p^1(0)}e^{\int_0^{s} 4\pi \gamma^* e^{\mu+\lambda}
(p+\rho-2j)d\eta}ds.
\end{equation}
For the interior integral expression we have since $p\leq \rho$,
\[
\int_0^{s} 4\pi \gamma^* e^{\mu+\lambda}
(p+\rho-2j)d\tau\leq \int_0^{s} 4\pi \gamma^* e^{\mu+\lambda}
(2\rho-2j)d\tau.
\]
Along the null geodesic we get
\[
\frac{d}{ds}\lambda(s,\gamma^*(s))=\lambda_t+\lambda_r \frac{d\gamma^*}{ds}=4\pi
\gamma^*(\rho-j)e^{\mu+\lambda}-\frac{m}{(\gamma^*)^2}e^{\mu+\lambda}.
\]
Hence by (\ref{lambdabound}), 
\begin{eqnarray*}
\int_0^{s} 4\pi \gamma^* e^{\mu+\lambda}
(2\rho-2j)d\tau 
&=&
2\lambda(s,\gamma^*(s))-2\lambda(0,\gamma^*(0))\nonumber\\
&&
{}+
\int_0^s \frac{2m(\tau,\gamma^*(\tau))}{(\gamma^*(\tau))^2}
e^{(\mu+\lambda)(\tau,\gamma^*(\tau))}d\tau \\
&\leq& 
C+C\int_0^s e^{\mu(\tau,\gamma^*(\tau))}d\tau.
\end{eqnarray*}
By \cite[Thm.~2.4]{AKR1} there exist positive constants 
$\alpha$ and $\beta$ such that if $(t,r)$ satisfies 
\[
r\geq 2M+\alpha e^{-\beta t}:=\sigma(t),
\]
then there is vacumm at $(t,r)$. 
By monotonicity of $\mu$ and the fact that there is vacumm for $r\geq\sigma$,
\[
 e^{\mu(\tau,\gamma^*(\tau))}
\leq  
e^{\mu(\tau,\sigma(\tau))}
\leq
e^{-\int_{\sigma(\tau)}^{\infty}\frac{M}{\eta(\eta-2M)}d\eta}
\leq 
C e^{-\frac{\beta\tau}{2}},
\]
and thus
\[
\int_0^{s} 4\pi \gamma^* e^{\mu+\lambda}
(2\rho-2j)d\tau\leq C.
\]
From (\ref{intt}) we therefore obtain the estimate
\[
\int_0^t\frac{ds}{p^0}
\leq 
C\int_0^t e^{(\mu-\lambda)(s,\gamma^*(s))}ds
=
C\int_0^t \frac{d \gamma^\ast (s)}{ds} ds 
< 2 C M,
\]
which says that $\gamma^\ast$ is incomplete, and the proof is complete. \prfe
\noindent
{\bf Acknowledgement\,:} The authors want to thank Helmut Friedrich whose question 
``Where is the matter?'' was a starting point for this work, and Mihalis Dafermos 
for discussions.

\end{document}